\documentclass[prb,aps,twocolumn]{revtex4}
\usepackage{graphicx}
\usepackage{dcolumn}
\usepackage{amsmath}
\usepackage{amssymb}
\def\be{\begin{equation}}
\def\ee{\end{equation}}

\def\ber{\begin{eqnarray}}
\def\eer{\end{eqnarray}}
\def\pv{{\bf p}}
\def\rv{{\bf r}}
\def\jv{{\bf j}}
\def\xv{{\bf {\rm x}}}
\def\Ev{{\bf E}}
\def\Av{{\bf A}}

\def\sigmat{\stackrel {\leftrightarrow}{\sigma}}
\def\rhot{\stackrel {\leftrightarrow}{\rho}}
\begin{document}
\title{Incompleteness of the Landauer Formula for Electronic Transport}
\author{Giovanni Vignale}
\affiliation{Department of Physics, University of Missouri-Columbia,
Columbia, Missouri 65211}
\author{Massimiliano Di Ventra}
\affiliation{University of California - San Diego, La Jolla, CA 92093}
\date{\today}
\begin{abstract}
We show that the Landauer multi-terminal formula for the conductance
of a nanoscale system is incomplete because it does not take into
account many-body effects which cannot be treated as contributions
to the single-particle transmission probabilities.  We show that the
physical origin of these effects is related to the viscous nature of the electron liquid,
and develop a perturbative
formalism, based on the time-dependent current-density-functional
theory, for calculating the corrections to the resistance in terms
of the ``Kohn-Sham current distribution'' and the
exchange-correlation kernel.  The difficulties that still remain in
calculating the latter are critically discussed.
\par
\end{abstract}
\maketitle
\section{Introduction}
The trend towards extreme miniaturization of electronic devices
provides strong motivation for theoretical studies aimed at
characterizing and understanding the electrical transport properties of
quantum-mechanical systems.~\cite{DiVentra2008} Here, by ``quantum system'' we mean a
molecular structure or cluster of atoms, or perhaps a microscopic region
defined on the surface of a semiconductor. Either way, this
system is connected to an external circuit which maintains
current flow via electron sources.

In the case of steady-state transport, this complicated
non-equilibrium many-body problem is often times simplified by {\em
conceptually} replacing the electron sources with ideal reservoirs,
whose role is to define a local electron distribution and a local
electrochemical potential at which electrons are injected in, or
extracted from the system.~\cite{Landauer57,landJCP89,Butt1} The
reservoirs are conceptual constructs which allow us to map the
transport problem onto an {\em ideal} stationary scattering one, so
that the time derivative of all local physical properties of the
system and the current, is zero.~\cite{DiVentraTod2004}

As a further simplification, one assumes that these reservoirs are
adiabatically ``connected'' to leads in which non-interacting
electrons are free to propagate before scattering at the lead-system
interface.~\cite{landJCP89} The leads are only a convenient region
of space where scattering states can be developed into an
appropriate basis of the Hilbert space. This viewpoint to electrical
conduction is known as Landauer approach.

\begin{figure}
\includegraphics[width=2.5in]{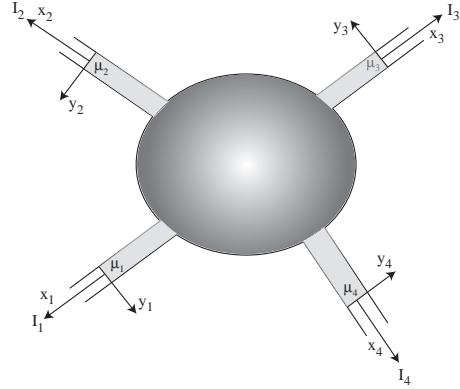}
\caption{Schematic of a quantum system in a multi-terminal
configuration. The central region is connected to leads of
non-interacting electrons, in turn connected adiabatically to
reservoirs of electrons.}\label{MesoCircuit}
\end{figure}

A schematic of this approach applied to a system connected to
several leads is shown in Fig.~\ref{MesoCircuit} where the shaded
region represents the system and the white regions are the leads,
numbered $1$ to $N$.
The contacts between the leads and the system can be very
complicated,  and should be considered  part of the system. The
proper lead, far from the contact, is a single-electron wave guide,
which we can assume to have constant electrochemical potential
$\mu_i$ (i=1,..,N).  At equilibrium all the leads are at the same
electrochemical potential $\mu$, and no current flows in or out of
the system. As we move slightly away from equilibrium, the currents
flowing in the leads will be related to the electrochemical
potentials by the linear relationships \be\label{LBFormula} I_i   =
\sum_{j=1}^N G_{ij}\mu_j~, \ee where the currents $I_i$ are reckoned
positive when they flow out of the system, and negative when they
flow into the system.  The coefficients $G_{ij}$ are the linear
conductances of the system. In an ideal steady-state situation
($\mu_i$ and $I_i$ independent of time) the conservation of charge
implies that the sum of all the currents is zero and therefore (for
every terminal $j$) \be \label{r1} \sum_{i=1}^N G_{ij} = 0~. \ee
Furthermore, the condition that the currents vanish when all the
chemical potentials are equal implies that \be \label{r2}
\sum_{j=1}^N G_{ij} = 0~. \ee It follows that the off-diagonal
conductances $G_{ij}$, with $i\neq j$ are sufficient to completely
characterize the linear steady-state response of the system.

The Landauer multi-terminal formula offers an appealing way to
relate $G_{ij}$ to the quantum-mechanical properties of the system.
In this theory $G_{ij}$ is proportional to the quantum-mechanical
probability that a single electron coming from lead $i$
with energy $E$ be transmitted into a different lead $j$ at the same
energy.  In linear response and at zero temperature, this energy can
be taken to be the Fermi energy $E_F$ of the system. We call this
coefficient $T_{ij}(E_F)$,  and notice that, in general, it is a sum
of all the partial probabilities of transmission from one of the
momentum states of the incoming electron at energy $E_F$ to one of
the momentum states of the outgoing electron at the same energy
(see, e.g., Ref.~\onlinecite{DiVentra2008}).  Thus the Landauer
formula reads \be\label{LBformula} G_{ij} = \frac{2e^2}{h}
T_{ij}(E_F),~~~~~~(i \neq j) \ee

It is important to note that the mathematical description of this
approach relies on scattering theory, namely on the transmission
properties of single electrons in the leads that scatter at the
leads-system interface. Therefore, for this description to be valid
{\em any interaction between electrons can only be included at a
mean-field level}. Many-body interactions beyond mean field destroy
the concept of single-particle transmission probability and, in
fact, when taken into account also in the leads, they do not even
allow for the derivation of a closed form for the total
current.~\cite{DiVentra2008,Meir,footnote0}
All of the above issues are particularly relevant in nanoscale
systems, where the current densities at the junction can be substantially larger
than in the bulk. A large current density implies a large number of
scattering events per unit time and unit volume, thus making the
description of transport phenomena in terms of non-interacting
particle properties questionable.~\cite{DiVentra2008}

It should thus not come as a surprise, and this is what we set to
clearly show in this paper, that {\em the Landauer
formula~(\ref{LBformula}) represents an incomplete description of
electrical transport in nanoscale systems}. This point is
particularly relevant these days, since there has been a surge of
theoretical activities aimed at calculating the transport properties
of these systems from ``first principles''. A popular way to tackle
this problem is to extract the transmission function appearing in
Eq.~(\ref{LBFormula}) from the one-electron Green's
function,~\cite{prec1} which is calculated from the self-consistent
potential of the ground-state density-functional theory
(DFT).~\cite{MDlong}  In this manner, one hopes to include the most
important effects of the electron-electron interaction without
losing the simplicity of the single-particle theory. Indeed, one
expects that interactions control the positioning of the
single-particle energy levels of the system with respect to the
Fermi level, and for this reason they have a large impact on the
conductance.

However, from a more general theoretical standpoint things are not
so simple.  First of all, even if we assume that the physical
approximations underlying the Landauer formula~(\ref{LBFormula}) are
a reasonable starting point to describe electrical transport, the
use of ground-state DFT in the present context is highly
questionable, since one effectively uses a ground-state theory for
an intrinsically non-equilibrium problem, even in linear response
and in the dc limit.~\cite{DiVentra2008,DiVentraTod2004}

Indeed -- and this leads us to the central message of this paper --
it is precisely the non-equilibrium nature of the transport problem
which renders Eq.~(\ref{LBformula}) untenable. In a practical
realization of a transport experiment, electrons are in a state of
non-equilibrium and, therefore, their correlations are time
dependent, even in the limit of zero frequency. These correlations
give rise to scattering processes that {\em cannot} be described by
a mean-field theory and, under certain conditions, may influence
substantially their dynamics.

Therefore, the Landauer formula, which has been derived within a
single-particle framework, cannot be uncritically transferred to the
many-body context, hoping that a proper inclusion of many-body
effects in the single-particle energy levels will always suffice. In
fact, in this paper we show that there are many-body corrections to
the Landauer formula, which {\it cannot} be formulated in terms of
single-particle transmission probabilities.

In order to demonstrate this important point of principle we start from the rigorous formulation of the conductance in terms of the zero-frequency limit of the exact non-local conductivity tensor $\sigma_{ij}(\rv,\rv';\omega)$ of the interacting many-electron system ($\omega$ is the frequency), in the linear
response regime. We then resort to the time-dependent current-density functional theory (TDCDFT)~\cite{Ghosh,VK,VUC} to show that the conductivity tensor $\sigmat$  satisfies the integral equation
 \be \label{sigmaDyson}
\sigmat = \sigmat_s- \sigmat \cdot \rhot_{xc}\cdot \sigmat_s \ee
where $\sigmat_s$ is the resistivity tensor of a noninteracting
system in the presence of a static potential $V_s$  (also known as
the Kohn-Sham potential) that reproduces the exact ground-state
density, and $\rhot_{xc}$  is a dynamical contribution that will be
defined precisely in Section~\ref{section3}.

The linear response formulation of mesoscopic transport dates back to works by Fisher  and Lee~\cite{fisher} and Baranger and Stone~\cite{Baranger} in the 1980s and has recently been combined with density functional theory by several authors.~\cite{Kamenev01,Koentopp,Bokes07,Prodan07}  Because $\sigmat_s$ is the conductivity of a noninteracting system,
it is possible to analyze it microscopically by the method of Fisher
and Lee~\cite{fisher} (later generalized by Baranger and
Stone~\cite{Baranger}), and thus show that this part of the
conductivity alone leads to the Landauer  formula~(\ref{LBFormula}),
with transmission probabilities computed from the Kohn-Sham
potential $V_s$. This step is still within the assumptions of the
Landauer approach, whereby the electron sources are replaced by
conceptual reservoirs whose role is to populate the single-particle
states according to different Fermi functions, and these states can
be developed in terms of the single-particle states of the leads.
However, this is not the whole story, since there is also the
contribution of the second term on the right hand side of
Eq.~(\ref{sigmaDyson}). In other words, within the Landauer viewpoint to conduction, even if
we knew the {\it exact} Kohn-Sham potential, including all the
self-interaction and non-local corrections which are responsible for
the correct alignment of the one-electron energy levels, we would
still be making an error in calculating the conductance from the
Landauer formula~(\ref{LBformula}).~\cite{prec2}

Next, we examine the nature of the correction to the Landauer
formula.  We observe that Eq.~(\ref{sigmaDyson}) is algebraically
equivalent to the equation
\be\label{rhoDyson} \rhot = \rhot_s +
\rhot_{xc} \ee where $\rhot \equiv \sigmat^{-1}$ is the exact
non-local resistivity, $\rhot_s \equiv \sigmat_s^{-1}$ is the
Kohn-Sham resistivity, and $\rhot_{xc}$ is the contribution from
many-body exchange and correlation. The resistivity controls the energy dissipation associated with a
steady current distribution, and the presence of the $xc$ correction
$\rhot_{xc}$ implies that there are mechanisms of dissipation that
are not taken into account in the Landauer approach of elastically
scattering electrons, with relaxation and dephasing occurring only
in the reservoirs.  What Eq.~(\ref{rhoDyson}) tells us is that
electron-electron interactions make up for additional dissipation
within the system, a dissipation that is physically a manifestation
of {\it electronic viscosity}. As a matter of fact, the simplest approximation for $\rhot_{xc}$,
which is derived from the Vignale-Kohn approximation to
TDCDFT,~\cite{VK} is expressed precisely in terms of the viscosity
of a homogeneous electron liquid: this approximation shows that
$\rho_{xc}$ is a positive kernel, always  giving rise to a positive
contribution to dissipation (i.e. an increase in resistance).

The existence of viscosity contributions to the electrical
resistance was first pointed out in Ref.~\onlinecite{Sai2005}, where
these contributions were called {\it dynamical corrections} because,
as we have discussed above and will show below, they vanish in a
strictly ground-state formulation of the theory.  However, the
relation of such contributions to the Landauer formula had remained
somewhat unclear (see also Ref.~\onlinecite{Koentopp}).  The present
work shows conclusively that the Landauer formula~(\ref{LBformula})
is incomplete, and the many-body corrections to it are precisely the
``dynamical corrections'' identified in Ref.~\onlinecite{Sai2005}.

The form of Eq.~(\ref{rhoDyson}) suggests a simple perturbative
approach to the calculation of the resistances $R_{ij}$  (derived
from the conductances $G_{ij}$ and defined more precisely below),
based on the minimal entropy-production principle of linear-response
theory.~\cite{Prigogine}  In brief, since the energy dissipation
rate (proportional to the entropy production) computed from the
single-particle (mean-field) theory is stationary with respect to a
small variation of the Kohn-Sham current distribution $\jv_s(\rv)$
(for given {\em total} currents in the leads) it follows that the
additional dissipation due to the $xc$ term is simply \be W_{xc} =
\jv_s \cdot \rhot_{xc} \cdot \jv_s~, \ee to first order in
$\rhot_{xc}$.  From this formula, and from the knowledge of the
Kohn-Sham current distribution, we can straightforwardly extract the
$xc$ contribution to the resistances. The formula for the $xc$
two-probe resistance of a quantum point contact or molecular
junction which we presented in Ref.~\onlinecite{Sai2005} will be
recovered as a special case of the general perturbative formulation.

Finally, we consider some quantitative aspects of the theory.  It
must be said that a compelling comparison between theory and
experiments is still hampered in most cases by an imperfect
characterization of the contact region.  Keeping this in mind, it is
now accepted that the theoretical calculations of the conductance of
molecular junctions, using the Landauer approach and ground-state
DFT, {\it overestimate} the measured conductance by at least an
order of magnitude.~\cite{MDlong}  Part of this discrepancy can
certainly be attributed to errors in determining the position of the
energy levels of the system relative to the electrochemical
potential in the leads -- errors which in turn are intimately
connected to self-interaction corrections, discontinuities in the
$xc$ potential as a function of particle number, and so
on.~\cite{Burke} Even after correcting for these effects, however,
it seems that the computed conductance remains larger than the
observed one, and it is here that our many-body corrections can play
a decisive role.

Our preliminary estimates of the size of the correction seem to indicate that the many-body viscous effects contribute only a small percentage to the total
resistance.~\cite{Sai2005} For the case of two infinite jellium electrodes separated by a vacuum
gap, the use of the viscosity as reported in Ref.~\onlinecite{CV} has shown an even smaller effect.~\cite{Jung}
But this does not mean that the issue is settled.

First of all, it is important to note that these estimates have been
based on an oversimplified description of the current density in
nanoscale systems, by neglecting transverse variations of both the
density and current density~\cite{Sai2005}. For instance, as shown
in Ref.~\onlinecite{Sai2007} transverse density gradients increase
the dynamical resistance. Quite generally, the transverse density
and current density gradients and the spatial variation of the
viscosity must all be taken into account when evaluating the viscous
resistance. This is particularly relevant in nanoscale systems where
non-linear (turbulent) effects have been recently
predicted.~\cite{DD,SBHD,BPD,BGD} Therefore, for a given nanoscale
system, these dynamical effects need to be evaluated with the
self-consistent microscopic density and current density
distribution.

Aside from the above issues, there remains another and more fundamental source of uncertainty -- namely, the value of the electronic viscosity which enters the dissipative kernel $\rho_{xc}$.  In the concluding part of this paper we will argue that this value is still subject to a large uncertainty and we will outline the path along which better approximations might be obtained.

This paper is organized as follows: Section~\ref{section2}  reviews
the general formulation for the conductance and the resistance of a
nanoscale system in terms of nonlocal conductivity. In
Section~\ref{section3} we  present the time-dependent current
density functional approach to the calculation of the resistivity
and demonstrate the existence of corrections to the Landauer
formula~(\ref{LBformula}). In Section~\ref{section4} we develop the
perturbative approach to the calculation of the many-body
corrections to the resistance. In Section~\ref{section5} we
illustrate the working of the formalism in a simple one-dimensional
model, re-deriving and extending the informal estimates of
Ref.~\onlinecite{Sai2005}. Finally, in Section~\ref{section6} we
discuss the present difficulties in performing accurate calculations
of the many-body corrections, and outline a path toward more
accurate estimates.

\section{Formulation}\label{section2}

Our starting point is the linear response formula for the steady
current density $\jv$ in the presence of a steady electric field
$\Ev$: \be \label{j-response} j_\alpha(\rv) = \sum_{\beta}\int
d\rv'~ \sigma_{\alpha\beta}(\rv,\rv')E_\beta(\rv') \ee where
$\alpha$ and $\beta$ denote cartesian indices and
$\sigma_{\alpha\beta}(\rv,\rv')$ is the real part of the
conductivity tensor. The integral runs over the whole volume of the
system depicted in Fig.~\ref{MesoCircuit}, including the leads. The
electric field, however, vanishes deep inside the leads. The above
equation is satisfied for small electric fields of the form \be
\Ev(\rv) = -\nabla_{\rv} \phi(\rv)\, \ee where $\phi(\rv)$ is an
electrostatic potential of arbitrary shape, except for the
constraint of tending to constant values \be
\label{chemicalpotentials} \phi (\rv) \to  \mu_i \ee deep into the
$i$-th lead.

Here, we assume that the electrostatic potential coincides with the
electrochemical potential deep into the leads.~\cite{precel}
Because a steady current also satisfies the continuity equation \be
\label{continuity} \nabla_{\rv} \cdot \jv(\rv)=0 \ee and because the
current cannot be affected by a uniform shift of the electric
potential in the whole space it follows that the conductivity tensor
satisfies the conditions\cite{Baranger} \be \label{c1}
\sum_{\alpha\beta}\partial_\alpha\partial_\beta^\prime
\sigma_{\alpha\beta}(\rv,\rv')=0\, \ee and \be \label{c2}
\sum_{\alpha\beta}\partial_\alpha \int_{C_n}  d
y_j'~\sigma_{\alpha\beta}(\rv,\rv')~\xv_{j\beta}= 0 \ee where
$\partial_\alpha$ is a short-hand notation for
$\frac{\partial}{\partial r_\alpha}$, and $\partial_\beta^\prime$
stands for $\frac{\partial}{\partial r_\alpha^\prime}$, $\hat \xv_j$
is the outwardly directed unit vector in lead $j$, and $y_j$ is a
short-hand notation for the coordinates perpendicular to
$\hat\xv_j$, which are integrated over the cross section $C_j$ of
the $j$-th lead (see Fig.~\ref{MesoCircuit} for a
schematic).~\cite{footnote1}

The current $I_i$ in the $i$-th lead is given by
\ber
I_i &=& \int_{C_i} dy_i~ \jv(\rv) \cdot \hat\xv_i \nonumber\\
&=& -\int_{C_i} dy_i~\int d\rv'  \sum_{\alpha\beta}\xv_{i\alpha} \sigma_{\alpha\beta}(\rv,\rv') \partial_\beta^\prime \phi(\rv') \nonumber\\
\eer Following Baranger and Stone~\cite{Baranger} we make use of
Eqs.~(\ref{chemicalpotentials}) and~(\ref{c1}), and an integration
by parts to find the intuitive result (cf. Eq.~(\ref{LBFormula}))
\be I_i = \sum_{j=1}^N G_{ij}\mu_j \ee where \be\label{BSformula}
G_{ij} = -\int_{C_i} dy_i\int_{C_j} dy_j^\prime
\sum_{\alpha\beta}\xv_{i\alpha} \sigma_{\alpha\beta}(\rv,\rv')
\xv_{j\beta}\,. \ee

Note that up to this point we have made no approximation on the
microscopic physical mechanisms that contribute to the
conductance~(\ref{BSformula}) apart from those embodied in the
viewpoint represented in Fig.~\ref{MesoCircuit}. Therefore, within
this viewpoint, the conductance~(\ref{BSformula}) contains, in
principle, all many-body interactions, even beyond mean field.

The next step is then to express the conductivity tensor in terms of
a microscopic current-current response function.  To this end we
introduce the {\it proper}  current-current response function, which
yields the electric current response to the fully {\em screened}
vector potential in the following manner \be \label{current-exact}
j_{\alpha}(\rv,\omega) = -e^2 \sum_\beta \int d\rv' \tilde
\chi_{\alpha\beta}(\rv,\rv';\omega)\left[A_\beta(\rv')+
A_{H,\beta}(\rv')\right]\,, \ee where $\bf A(\rv)$ is the external
vector potential and ${\bf A}_H(\rv)$ is the vector potential
additionally created by the screening charge.\cite{footnote2} The
factor $e^2$ ($e$ being the absolute value of the electron charge)
is introduced to be consistent with the definitions used in other
publications~\cite{DiVentra2008,VUC,GV05}. Then the conductivity is
\be\label{standard} \sigma_{\alpha\beta}(\rv,\rv') =
-e^2\lim_{\omega \to 0}  \frac{\Im m \tilde \chi_{\alpha
\beta}(\rv,\rv';\omega)}{\omega}\,. \ee

\begin{figure}
\includegraphics[width=3.0in]{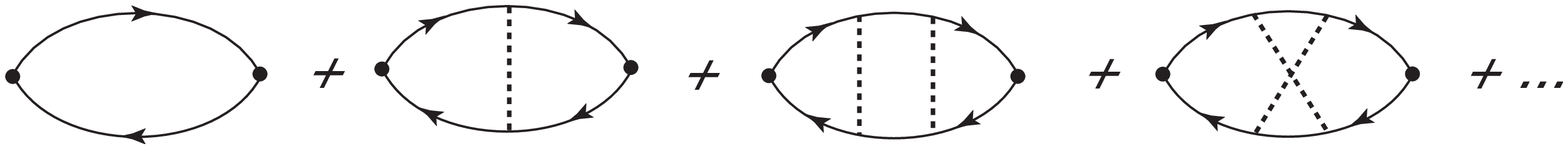}
\caption{Diagrams for the proper current-current response function.
The solid dots represent (particle) current vertices. The solid
lines represent the free particle propagator and the dotted lines
Coulomb scattering processes.}\label{Diagrams}
\end{figure}

The proper current-current response function is best expressed in
terms of an infinite series diagrams with two current vertices, such
as the diagrams shown in Fig.~\ref{Diagrams} where the solid lines
represent free particle propagation and dotted lines Coulomb
interactions. Notice that this series does not contain any diagrams
that can be divided into two parts by cutting a single Coulomb
interaction line. It is the exclusion of these diagrams that makes
our response function ``proper'', as opposed to
``full''.\cite{footnote2}

\subsection{Mean-field approximation}

In the special case of a non-interacting system, or a system
interacting at a mean-field level, only the first term of the series
survives and we get, following the standard rules~\cite{GV05}
\be\label{sigmanoninteracting} \sigma_{\alpha\beta}(\rv,\rv') = -\pi
e^2\sum_{nm}\frac{\partial f(\epsilon_n)}{\partial \epsilon_n}
W^{\alpha *}_{nm}(\rv) W^{\beta}_{mn}(\rv^\prime)
\delta(\epsilon_n-\epsilon_m)~, \ee where $n$ and $m$ denote exact
single-particle eigenstates with energies $\epsilon_n$ and wave
functions $\psi_n(\rv)$, $f(\epsilon_n)$ is the Fermi distribution
at the common chemical potential $\mu$ (before applying the bias)
and temperature $T$, and $W^{\alpha}_{nm}(\rv)$ is the matrix
element of the $\alpha$-component of the (particle) current operator
between states $m$ and $n$: \be W^{\alpha}_{nm}(\rv) =
-\frac{i\hbar}{m}\left\{\psi_n^*(\rv)\partial_\alpha \psi_m(\rv) -
[\partial_\alpha\psi_n^*(\rv)] \psi_m(\rv) \right\}\,. \ee

Eq.~(\ref{sigmanoninteracting}) leads upon substitution in
Eq.~(\ref{BSformula}) to the standard Landauer
formula~(\ref{LBformula}).~\cite{fisher,Baranger} The calculation is
quite subtle, hinging on the possibility of choosing a complete set
of exact eigenstates in the form of scattering states, i.e. states
of energy $\epsilon$ which describe a single particle ``entering''
the system in the transverse channel $a$ of the $i$-the lead,  and
scattered with probability amplitude $t_{ia,jb}$ into any transverse
channel $b$ of the $j$-th lead.

Within this mathematical assumption, the transmission coefficient
$T_{ij}$ that appears in Eq.~(\ref{LBformula}) is found to be given
by \be T_{ij} = \sum_{a,b}\vert t_{ia,jb}\vert^2~. \ee We refer to
the original papers~\cite{fisher,Baranger} for the details of this
derivation. What is important for our purposes is that the
conventional Landauer multi-terminal formula~(\ref{LBformula})
emerges from an approximation to the exact formula (\ref{BSformula})
-- an approximation in which only the first term in the infinite
series of diagrams for the proper current-current response function
is retained.

The question now arises how to go beyond this simplest approximation
to include electron-electron interaction effects.  In the next
section we describe an approach based on time-dependent
current-density functional theory.

\section{Time-dependent current-density functional theory}\label{section3}

As discussed in the Introduction, a popular approach to the
inclusion of many-body effects in nanoscopic transport is to use the
Landauer formula (\ref{LBformula}), but calculate the transmission
probabilities by solving the one-particle scattering problem in a
static effective potential that includes many-body effects.  How is
such a potential to be constructed?

The ground-state density functional theory (DFT) of Hohenberg, Kohn
and Sham offers a practical answer.~\cite{HKS}  According to this
theory it is possible to find, in principle, an exchange-correlation
potential which, in combination with the Hartree potential and the
external potential, produces the correct ground-state density of the
many-body system. Furthermore, this potential (known as the
Kohn-Sham potential) is uniquely determined by the density.
Thus, it is very tempting to make use of the Kohn-Sham potential to
calculate the transmission probabilities and hope that all many-body
effects pertaining to the transport problem be included.
Unfortunately this approach lacks any rigorous theoretical
foundation. In practice, it amounts to ``dressing up'' the free
particle lines in the first diagram of Fig.~\ref{Diagrams}, while
still discarding all the other diagrams. Therefore, it must be
interpreted as nothing more than a single-particle {\em mean-field}
approximation even if we knew the {\em exact} ground-state $xc$
functional, and as such there is no physical reason why this should
be even approximately correct.

The time-dependent current density functional theory offers a more solid basis to attack the problem.  Taking for granted the ordinary DFT description of the ground-state, the TDCDFT attempts to describe the current response of the many-body system as the response of a non-interacting reference system to an effective time-dependent vector potential.  The non-interacting reference system is usually taken to be the ``Kohn-Sham system", i.e. the non-interacting system that is used in ordinary DFT to reproduce the ground-state density of the many-body system.  Thus, in the TDCDFT approach the current response to a time-periodic vector potential $\Av(\rv,t) = \Av(\rv,\omega)e^{-i\omega t}+ c.c$, is written as
\ber\label{current-CDFT}
j_\alpha(\rv, \omega) &=& -e^2\sum_\beta \int d\rv' \chi_{s,\alpha\beta}(\rv,\rv';\omega)\left\{A_\beta(\rv',\omega)\right.\nonumber\\
&+&\left.A_{H,\beta}(\rv',\omega)+A_{xc,\beta}(\rv',\omega)\right\}
\eer where $\chi_{s,\alpha\beta}(\rv,\rv';\omega)$ is the current
response function of the Kohn-Sham system, $\Av_H$ is the Hartree
vector potential, and $\Av_{xc}$ is the exchange-correlation vector
potential.\cite{footnote3} The essential point is that the
exchange-correlation potential is a unique functional of the current
density, and in the linear approximation can be represented as
\be\label{kernel} -e^2A_{xc,\alpha}(\rv,\omega)=\int d\rv^\prime
\sum_\beta f_{xc,\alpha\beta}(\rv,\rv';\omega)
j_\beta(\rv',\omega)\,, \ee where the ``exchange-correlation kernel"
$f_{xc,\alpha\beta}(\rv,\rv';\omega)$ is determined by the
ground-state density.  Later we will also need the
``exchange-correlation electric field'', which is defined as
\be\label{Def-Exc} E_{xc,\alpha}(\rv,\omega)= i \omega
A_{xc,\alpha}(\rv,\omega)\,. \ee

The above formula~(\ref{current-CDFT})  should be compared with the exact linear response formula~(\ref{current-exact}).
Combining Eqs.~(\ref{current-CDFT}), (\ref{kernel})
and~(\ref{current-exact}) we find the well-known relation between
$\tilde \chi$, $\chi_{s}$, and $f_{xc}$, namely \be \label{chiinv}
[\tilde \chi^{-1}]_{\alpha\beta}(\rv,\rv',\omega) =
[\chi_s]_{\alpha\beta}^{-1}(\rv,\rv',\omega) -
f_{xc,\alpha\beta}(\rv,\rv',\omega) \ee where $[\tilde
\chi^{-1}]_{\alpha\beta}(\rv,\rv',\omega) $ is the matrix inverse of
$\tilde \chi_{\alpha\beta}(\rv,\rv',\omega)$, which is regarded as a
matrix with indices $\alpha,\rv$ and $\beta,\rv^\prime$.

Eq.~(\ref{chiinv}) gives us a handle on the inverse of the
conductivity, i.e., the resistivity tensor.  To make the connection,
observe that the complex conductivity tensor $\tilde
\sigma (\rv,\rv',\omega)$  is given by \be\label{standard2} \tilde
\sigma_{\alpha\beta}(\rv,\rv',\omega) =-e^2\frac{\tilde
\chi_{\alpha\beta}(\rv,\rv';\omega)}{i\omega}\,, \ee whose real part
reduces to  $\sigma (\rv,\rv')$ in the limit $\omega \to 0$.
Accordingly, the complex resistivity tensor is given by \be \tilde
\rho_{\alpha\beta}(\rv,\rv',\omega)\equiv[\tilde
\sigma^{-1}]_{\alpha\beta}(\rv,\rv',\omega) =-\frac{i\omega}{e^2}
[\tilde \chi^{-1}]_{\alpha\beta}(\rv,\rv',\omega). \ee Then making
use of Eq.~(\ref{chiinv}) we find \be \tilde
\rho_{\alpha\beta}(\rv,\rv',\omega)=\tilde
\rho_{s,\alpha\beta}(\rv,\rv',\omega)+ \frac{i\omega}{e^2}
f_{xc,\alpha\beta}(\rv,\rv',\omega)\, \ee where the ``Kohn-Sham
resistivity", $\tilde \rho_s$, has the same relation to $\chi_s$ as
the full interacting resistivity to $\tilde \chi$. Finally, taking
the real part of both sides and going to the zero-frequency limit we
find \be\label{sigmaDysonbis} \rho_{\alpha\beta}(\rv,\rv') =
\rho_{s,\alpha\beta}(\rv,\rv') +\rho_{xc,\alpha\beta}(\rv,\rv')\,
\ee where \be\label{rhoxc} \rho_{xc,\alpha\beta}(\rv,\rv') \equiv
-\lim_{\omega \to 0} \frac{\omega}{e^2}\Im m
f_{xc,\alpha\beta}(\rv,\rv',\omega)~. \ee

This is the main result of our paper which we have anticipated in
Eq.~(\ref{rhoDyson}).  We clearly see that the resistivity of the
Kohn-Sham system -- a non-interacting system in which many-body
effects enter only implicitly through the static
exchange-correlation potential -- is not the whole story. This means, in particular, that {\em it is not possible to give an
exact representation of the conductance in terms of single-particle
transmission probabilities}. We have demonstrated this point for the
linear-response regime, namely in the limit of zero external bias.
However, this result must be valid also out of linear response, even
though in this case the extent of the dynamical corrections, which
in linear response are embodied in $f_{xc,\alpha\beta}$, is not so
easily determined.

We now shift our attention to the estimate of the dynamical
exchange-correlation contribution $\rho_{xc}$ which is controlled
entirely and explicitly by many-body effects, i.e., time-dependent
correlations in the effective potential of TDCDFT.  The many-body kernel $f_{xc}$, which appears in Eq.~(\ref{sigmaDysonbis}), is not known exactly for any system, but a local
approximation to it is available and has been used in the recent
literature with varying degrees of
success.~\cite{VK,Vignale06,Sai2005,Jung,Sai2007} In a local approximation the
key quantity \be\label{Exc} \tilde E_{xc,\alpha}(\rv,\omega)
=-\frac{i\omega}{e^2} \sum_\beta\int d \rv^\prime
f_{xc,\alpha\beta}(\rv,\rv',\omega) j_\beta(\rv^\prime,\omega)~, \ee
which has the physical significance of ``exchange-correlation
electric field" (see Eq.~(\ref{Def-Exc})), is taken to be a function
of the local value of $\jv(\rv)$ and its first and second spatial derivatives.
The simplest approximation
in this class is the so-called adiabatic local density approximation
(ALDA), which provides an instantaneous connection between $\tilde
\Ev_{xc}(\rv)$ and $\jv(\rv)$.  In this approximation, however,
$f_{xc}$ is purely real, resulting in an exchange-correlation
electric field that is always $90^0$ out of phase with the current
density. Therefore, such a field cannot contribute to the d.c.
resistivity, consistent with the fact that $\Im m f_{xc}$ vanishes
in this approximation.

So in order to obtain exchange-correlation corrections to the
resistivity we must go beyond the adiabatic approximation.  This can
be done with the help of the VK local density
approximation,\cite{VK} which calls into play the viscosity of the
electron gas.   In this approximation the exchange-correlation field
has a dissipative component which is 180$^0$ out of phase, i.e.,
opposite to the current.  In the zero-frequency limit this component
of the exchange-correlation field has the form~\cite{VUC}
\ber\label{VKExc}
&&E_{xc,\alpha}(\rv)  = -\sum_\beta \int d \rv^\prime \rho_{xc,\alpha\beta}(\rv,\rv')j_{\beta}(\rv')\nonumber\\
&\simeq& \frac{1}{e^2n(\rv)}\sum_\beta\partial_\beta \left\{\eta (\rv)\left[\partial_\beta\left(\frac{j_\alpha(\rv)}{n(\rv)} \right)+\partial_\alpha\left(\frac{j_\beta(\rv)}{n(\rv)} \right) \right.\right.\nonumber\\
&-&\left.\left.\frac{2}{3}\nabla_{\rv}\cdot\left(\frac{\jv(\rv)}{n(\rv)}
\right) \delta_{\alpha\beta} \right]\right\}, \eer where $\jv(\rv)$
is the {\it electric} current density, and $\eta(\rv)$ is the d.c.
shear viscosity of a homogeneous electron gas of density
$n(\rv)$.\cite{CV}  Eqs.~(\ref{sigmaDysonbis}), (\ref{rhoxc})
and~(\ref{VKExc}) constitute a complete (albeit approximate)
formulation of the microscopic resistivity tensor within TDCDFT.  The
Kohn-Sham resistivity itself is accessible from the ordinary static
DFT. $\rho_{xc}$ is best described through the effective electric
field it produces -- an electric field $\Ev_{xc}$ directed {\em
against} the current, which therefore does negative work on the
current.   In the next section we show how our expression for
$\Ev_{xc}$ can be directly applied to the calculation of the
macroscopic conductance.

\section{Perturbative calculation of the conductance}\label{section4}
Let us return to the system shown in Fig.~\ref{MesoCircuit} and let
us assume that the electrochemical potentials $\mu_i$  are
periodically modulated in time with a (very) small angular frequency
$\omega$. By ``small'' we mean a frequency much smaller than any
other internal frequency of the system. The lead currents induced by
the modulation are then also periodic and given by Eq.~(\ref
{LBFormula}). Since the reservoirs are the only part of the system
on which we have direct control it is evident that the work done on
the system per unit time is \be\label{dissipation1} W =
\sum_{i=1}^N\langle I_i \mu_i \rangle \ee where the angular brackets
denote a time-average over a period of oscillation.  This is also
the energy that must be internally dissipated if the system is to
remain in the steady state.

In order to express $W$ in terms of the lead currents we must invert
the linear relation~(\ref{LBFormula}) between the lead currents and
the electrochemical potentials.  Strictly speaking, this relation is
not invertible, because a rigid shift of all the electrochemical
potentials has no effect on the current. But the problem is easily
solved by permanently grounding one of the reservoirs, say the one
with $i=1$ so that $\mu_1=0$ at all times.  Then the linear relation
between the remaining $N-1$ currents, $I_2, ..., I_{N}$ and the
corresponding electrochemical potentials  $\mu_2, ..., \mu_{N}$ {\it
is} invertible, and the current in the grounded lead is simply given
by $I_1=-I_2 -...-I_N$.

Then, we see that the dissipated power can be represented as
\be\label{dissipation2} W=\sum_{i,j=2}^N \langle I_i R_{ij}I_j
\rangle \ee where the $(N-1)\times (N-1)$ matrix $R_{ij}$ is the
inverse of the matrix $G_{ij}$ stripped of the first row and the first
column.\cite{footnote4}   The macroscopic expression for $W$ in
terms of $R_{ij}$  is now equated to the usual microscopic
expression in terms of the resistivity, resulting in the following
equation:
 \be\label{dissipation3}
\sum_{i,j=2}^N  I_i R_{ij}I_j  =   \int d\rv \int d\rv' \jv(\rv)
\cdot \rhot(\rv,\rv') \cdot \jv(\rv')\,, \ee where the integrals run
over the volume of the system, including the leads, and we have
dropped the time average by going to the zero-frequency limit.
Finally, by making use of Eq.~(\ref{sigmaDysonbis}) we arrive at
\be\label{dissipation4} \sum_{i,j=2}^N  I_i R_{ij}I_j  =   \int d\rv
\int d\rv' \jv(\rv) \cdot \left[\rhot_s(\rv,\rv')+\rhot_{xc}\right]
\cdot \jv(\rv')\,, \ee This equation is formally exact if we know
the exact many-body kernel $f_{xc}$. Let us compare it with the
formula we would obtain from the conventional single particle
theory, i.e. from the Landauer formula~(\ref{LBformula}), {\it for
the same lead currents}.

As discussed in Section~\ref{section2}, the ``single-particle"
theory assumes that all many-body effects can be included in the
Kohn-Sham potential of the ground-state.  Apart from this, the
system is non-interacting. The current density distribution
$\jv_s(\rv)$ of this fictitious Kohn-Sham system is in general
different from the true current density distribution $\jv(\rv)$ of
the many-body system, even though the macroscopic lead currents,
i.e. the fluxes of $\jv$ and $\jv_s$ into the leads, are imposed to
be the same. Therefore we write \be\label{dissipation5}
\sum_{i,j=2}^N I_i R_{s,ij}I_j  = \int d\rv \int d\rv' \jv_s(\rv)
\cdot \rhot_s(\rv,\rv')\cdot \jv_s(\rv')\,, \ee where $R_{s,ij}$ are
the macroscopic resistances of the Kohn-Sham system, obtained from
the standard single-particle theory.  Notice that the dynamical term
$\rhot_{xc}$ is absent in the single-particle theory.

The comparison between Eqs.~(\ref{dissipation4}) and
~(\ref{dissipation5}) is complicated, in general,  by the difference
between $\jv$ and $\jv_s$.  A simple comparison becomes possible in
the perturbative limit, i.e. under the assumption that the dynamical
many-body correction embodied in $\rhot_{xc}$ is small.  To this end
we observe two facts:  (i) The difference between $\jv$ and $\jv_s$
is of first order in the $xc$ correction, and (ii) the right hand
side of Eq.~(\ref{dissipation5}) is stationary under a small
variation of the current distribution, such as the difference
between $\jv$ and $\jv_s$.   The physical reason for this is that in
any linear system with external leads (such as the Kohn-Sham system
we are considering here) the power dissipated is stationary with
respect to an infinitesimal variation of the current distribution at
{\em constant} lead currents.  This implies that a first-order
variation in the current distribution (about the steady distribution
$\jv_s$ in this case) produces a second-order variation in the
dissipated power. Taking this into account we see that we can safely
replace $\jv_s$ by $\jv$ in Eq.~(\ref{dissipation5}) and subtracting
from Eq.~(\ref{dissipation4}) we arrive at the main result of this
section: \be \label{mainresult} \sum_{i,j=2}^N  I_i \Delta R_{ij}
I_j=  \int d\rv \int d\rv' \jv_s(\rv) \cdot
\rhot_{xc}(\rv,\rv')\cdot \jv_s(\rv')\,, \ee where \be \Delta R_{ij}
\equiv R_{ij}-R_{s,ij} \ee is the dynamical correction to the
resistance. This equation expresses the many-body correction to the
macroscopic resistance in terms of two things that are approximately
known and/or calculable, namely,  the resistivity
exchange-correlation kernel $\rhot_{xc}$, defined by
Eq.~(\ref{VKExc}), and the Kohn-Sham current distribution of the
device, $\jv_s(\rv)$, associated with the macroscopic lead currents
$I_i$. The latter can be calculated, in principle, from the response
of a non-interacting system to a screened electric field. In
practice, one can calculate the correction to $R_{ij}$ by
considering a special situation in which only $I_i$ and $I_j$ are
different from zero.  Then the left hand side of
Eq.~(\ref{mainresult}) gives us exactly the desired correction to
$R_{ij}$.

\section{An example}\label{section5}

\begin{figure}
\includegraphics[width=3.0in]{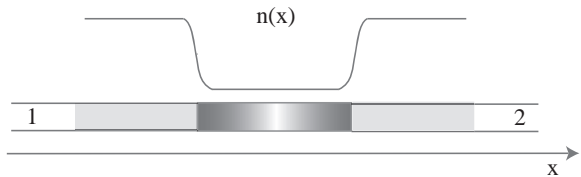}
\caption{A model two-terminal device in which the density changes only in one dimension. 1 and 2 are the lead regions.}\label{1DModel}
\end{figure}

Let us consider a simple model application of the general formalism.
Our system is a potential barrier connected by two
identical homogeneous leads (labeled 1 and 2 in Fig.~\ref{1DModel})
to two reservoirs, a ``source" and a ``drain", aligned along the $x$
axis. The system is perfectly homogeneous in the transverse
directions $y$ and $z$.  The density changes only in the $x$ direction. The source (terminal 1) is grounded, so we
only need to determine the two-terminal resistance $R_{22}$ or,
equivalently, the conductance $G_{22} =R_{22}^{-1}$. Let $I$ be the
current through the device and $\jv(x) =I/A$ the current density,
where $A$ is the transverse cross section of the device.

Notice that in this case there is no question of $\jv(x)$ being
different from $\jv_s(x)$ since, by continuity, they are both
uniform and equal to $I/V$.  In the absence of electron-electron
interactions the conductance of this system is simply given by
Eq.~(\ref{LBformula}) --  the total transmission probability across
the potential barrier being the sum of the transmission
probabilities of all the occupied transverse modes.  Including the
electron-electron interaction has two effects. The first is fairly
trivial, namely, the effective potential in the ground-state is
modified by screening and exchange-correlation effects, and the
transmission probabilities must be recalculated for this effective
potential. Up to this point the single-particle
formula~(\ref{LBformula}) remains in force.

The second effect is the dynamical exchange-correlation correction
-- an effect that cannot be forced into the mold of the static
mean-field theory. Making use of Eq.~(\ref{mainresult}) with $N=2$
and $j(x)=I/A$ we obtain
 \be\label{example1}
 \Delta R_{22} =   \frac{1}{A^2}\int d \rv \int d \rv' \rho_{xc}(x,x')\,
 \ee
where $\rho_{xc}(x,x')$ denotes the $xx$ component of the tensor $\rhot_{xc}(x,x')$, and we emphasize the fact that, due to our assumptions, it depends only on $x$ and $x'$.
Now observe that, according to Eq.~(\ref{VKExc}) the action of $\rho_{xc}$ on a uniform current density is specified by
\be\label{example2}
 \int d\rv' \rho_{xc}(x,x') = -\frac{4}{3e^2n(x)}\partial_x \left[\eta(x) \partial_x\frac{1}{n(x)}
 \right]\,,
 \ee
 where $\eta(x)$ is the shear viscosity of the homogenous electron gas evaluated at the ground-state density $n(x)$.
 Substituting this in Eq.~(\ref{example1}) and doing an integration by parts we arrive at
 \be\label{example3}
 \Delta R_{22} = \frac{4}{3 e^2 A} \int dx \eta (x)\frac{[n'(x)]^2}{[n(x)]^4}~,
 \ee
 where $n'(x)=\partial_xn(x)$. Notice that this has the correct dimensions (Ohm) because $n$ is a three-dimensional density and $\eta$ has the dimensions of $\hbar$ times a density.
 Fig.~\ref{1DModel} shows the electronic density in the leads and in the device.  Clearly the dynamical correction comes entirely from the non-homogeneous regions near the edges of the barrier (the contacts).

Eq.~(\ref{example3}) was first obtained in
Ref.~\onlinecite{Sai2005}, in a more intuitive manner.  The
advantage of the present formulation is that it allows easy
extension to more complicated situations. For example, we can
include the dependence of the density on the transverse coordinates
$y$ and $z$, while neglecting variations of the transverse
components of the current. In this case, we still have $\jv(\rv)$ =
constant, but now the gradient of the density has both longitudinal
and transverse components.  As a result we get \be\label{example4}
 \Delta R_{22} = \frac{1}{e^2A^2} \int d\rv~ \eta (\rv)\left\{\frac{4}{3}\frac{\vert \nabla_\parallel n(\rv)\vert^2}{[n(\rv)]^4}+\frac{\vert \vec \nabla_\perp n(\rv)\vert^2}{[n(\rv)]^4} \right\}~,
 \ee
where $\nabla_\parallel$  is the $x$-component of the gradient and $\vec \nabla_\perp$ is the gradient in the $y-z$ plane.  This result (for constant viscosity) was first reported in
Ref.~\onlinecite{Sai2007} following an intuitive procedure, still
based on the calculation of the power dissipated in the circuit.

\section {Discussion and critique}\label{section6}
In this paper we have shown that the single-particle mean-field
framework of Landauer is inadequate {\it in principle} to describe
the transport problem in nanoscale systems. That is to say a
calculation of conductance from Eq.~(\ref{LBformula}) would not
provide the exact current even if one could determine the
transmission probabilities with the utmost precision. Dynamical,
many-body effects enter the picture due to the {\em intrinsic}
non-equilibrium nature of conduction.~\cite{DiVentra2008} These
effects cannot be captured by a static formulation.

\begin{figure}
\includegraphics[width=3.0in]{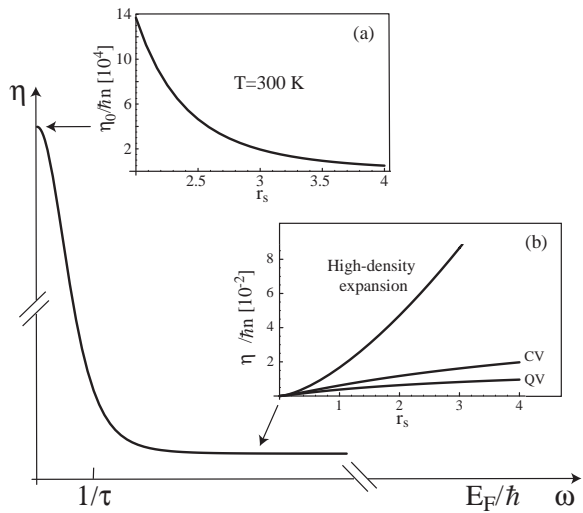}
\caption{Qualitative behavior of the viscosity of a homogeneous electron gas as a function of frequency.  For frequencies smaller than $1/\tau$, with $\tau$ the quasiparticle relaxation time, $\eta$ approaches the d.c. limit $\eta_0$. $\eta_0$ vs. density is shown in inset (a) for $T =300$ K. The calculation is done with the Abrikosov-Khalatnikov formula~\cite{AK}.   For frequencies larger than $1/\tau$, but still much smaller than the Fermi frequency $E_F/\hbar$,  $\eta$ tends to a different constant $\eta_\infty$. The behavior of $\eta_\infty$ vs density is shown in  inset (b)  in various approximations at zero temperature. CV is from Ref.~\cite{CV}, QV is from
Ref.~\onlinecite{QV}. The atomic unit of viscosity is $\hbar/a_B^3 =
\simeq 7 \times 10^{-3}$ Poise  ($a_B$ is the Bohr radius) and
$na_B^3 = \frac{3}{4 \pi r_s^3}$.}\label{Viscosity}
\end{figure}

The next question is: what is the actual size of these dynamical
corrections? One of the main results of this paper,
Eq.~(\ref{mainresult}), opens the way to a fully microscopic
first-principles calculation of nanoscopic resistances and
conductances (in the linear regime) within the framework of the
local approximation to time-dependent current density functional
theory. In Ref.~\onlinecite{Sai2005} we tried to address this question for
the simple quasi-one-dimensional model discussed in the previous
section and found that the viscosity correction to the resistance
was only a small fraction of the total. For the case of two infinite
jellium electrodes separated by a vacuum gap, a more accurate
calculation based on the homogeneous electron gas viscosity reported
in Ref.~\onlinecite{CV} (see Fig.~\ref{Viscosity}) has shown an even
smaller effect.~\cite{Jung} But, as pointed out in the introduction,
this does not mean that the issue is settled.

Looking back at Eq.~(\ref{mainresult}) we see that an accurate
evaluation of the viscosity correction has two ingredients: (1) the
Kohn-Sham current density distribution and (2) the viscosity of the
homogeneous electron gas.  As for the current density, it is
important to note that all the estimates so far have been based on
an oversimplified model in which the current density was assumed to
be uniform in space~\cite{Sai2005,Jung,Sai2007}.  In general, the
spatial variation of the current density  cannot be neglected;
especially in nanoscale  systems where large transverse variations
of the current density are common.~\cite{DD,SBHD,BPD,BGD}

Another and more fundamental source of uncertainty is in the value
of the electronic viscosity which enters the dissipative kernel
$\rho_{xc}$.   The viscosity we have used so far, which is plotted
in inset (b) of Fig.~\ref{Viscosity}, was obtained from a zero-temperature
calculation in the limit of zero frequency.  In other words, the
temperature (and, with it, the quasiparticle scattering rate) went
to zero {\it before} the frequency.  When calculated in this manner,
the viscosity turns out to be very small indeed: its value is in the
range of $10^{-5}/r_s^3$ Poise, where $r_s= \left(\frac{3}{4 \pi n
a_B^3}\right)^{1/3}\simeq 1$ is  the average inter-electron distance
in units of the Bohr radius $a_B$.  (For comparison, water at room
temperature has a viscosity of about 10$^{-2}$ Poise).

On the other hand, it is well known from the theory of homogeneous
Fermi liquids~\cite{AK}, that the behavior of the viscosity is
quite different if the zero-frequency limit is approached at finite
temperature.  Namely, in this case the viscosity turns out to be
proportional to the mean free path of the quasiparticles, which
grows as $1/T^2$ in the low temperature limit. The divergence of the
zero-frequency viscosity for $T \to 0$ reflects the fact that
long-lived quasiparticles can transport momentum arbitrarily far
away from the source of the stress.  This is also the reason why the
viscosity of an ideal classical gas is independent of density,\cite{Landau}
since an increase or a decrease in the frequency of molecular
collisions is exactly compensated by an opposite variation in the
molecular mean free path.

%

Panel (b) of Fig.~\ref{Viscosity} shows the behavior of the
zero-frequency viscosity at room temperature for an electron gas,
estimated from Eq. (7.22) of Abrikosov and Khalatnikov
(AK),~\cite{AK} with due allowance made for the different form of
the interaction potential (the AK work was for $^3$He).  The Eq.
(7.22) of AK can be rewritten as \ber\label{AK7.22}
\eta_0 &=&\hbar n \frac{8}{15 \pi} \left(\frac{E_F}{k_BT}\right)^2 (k_F a_B)^2 \nonumber \\
&\times&\left\{\left[\frac{\bar w(\theta,\phi)}{\cos\frac{\theta}{2}}(1-\cos \theta)^2\sin^2\phi
\right]_{av}\right\}^{-1}\,.
\eer
which shows explicitly the physical dimensions of the viscosity ($\hbar n$). Here $k_F$ is the Fermi wave vector, $E_F = \frac{\hbar^2 k_F^2}{2m}$ is the Fermi energy, $a_B$ is the Bohr radius and $\bar w(\theta,\phi)$ is the square of the matrix element of the electron-electron interaction potential (expressed in units of $4\pi e^2/k_F^2$)  between the initial and the final state of a collision process with incoming momenta $\pv_1,\pv_2$ and outgoing momenta $\pv'_1,\pv'_2$, where $\theta$ is the angle between the incoming momenta and $\phi$ is the angle between the planes formed by  $(\pv_1,\pv_2)$ and $(\pv'_1,\pv'_2)$.  All the momenta are close to the Fermi surface, and the symbol $av$   denotes the average over $\theta$ and $\phi$.   The simplest approximation for $\bar w$ is the Thomas-Fermi approximation, in which we have
\be
\bar w(\theta,\phi) = \left(\frac{1}{2(1-\cos \phi)+4 \alpha r_s/\pi}\right)^2\,,
\ee
with $\alpha =(4/9\pi)^{1/3} \simeq 0.521$.    With this approximation, the average over $\theta$ can be done analytically  and Eq.~(\ref{AK7.22}) can be rewritten as
\ber
\frac{\eta_0}{\hbar n} &=& \frac{1}{8 (\alpha r_s)^6}
\left(\frac{1.579 \times 10^5}{T}\right)^2  \nonumber\\
&\times&\left\{\int_0^\pi d\phi \frac{\sin^2 \phi}{\left(4 \sin^2
\frac{\phi}{2}+\frac{4 \alpha r_s}{\pi}\right)^2}\right\}^{-1} \eer
The result of the evaluation of this expression is shown in inset
(a) of Fig.~\ref{Viscosity} for $T=300$~K (room temperature). Notice
that the presence of the factor $1/r_s^6$ causes the viscosity to
increase sharply with increasing density, in  contrast with what we
observe in inset (b) of Fig.~\ref{Viscosity}. It is evident that the
d.c. viscosity is orders of magnitude larger than the finite
frequency viscosity plotted in inset (b) of Fig.~\ref{Viscosity}.

What does this imply for our analysis of the conductance in nanoscopic and mesoscopic  devices? Obviously, these systems do not host a uniform
electron liquid, and in particular they do not support
long-lived quasiparticles that can transport momentum to infinity.
This means that the large finite-temperature results of the uniform
electron liquid are almost certainly not relevant for nanoscopic
devices: the mean free path of quasiparticles is naturally limited
by the geometric size of the device.~\cite{DiVentra2008,DD1}
However, the huge difference between the numerical values of the
viscosities in the insets of Fig.~(\ref{Viscosity})
suggests the possibility of a mesoscopic ``middle-ground"  which
under certain conditions may be much larger than the
zero-temperature viscosity. This, however, is unlikely to be
``universal'', rather it must be related to the specific microscopic
geometry of the system.  A related difficulty is that, in general, the rate of dissipation in
an interacting electron system depends strongly on the excitation
spectrum of the system. Modeling dissipation through the viscosity
of a homogeneous electron gas, as implied by our local-density
approximation, may lead to a severely distorted description of the
dissipative process.  A truly universal description of dissipation
(if possible at all) is still out of sight.

A central issue emerges from the above discussion, namely the need
for an accurate, testable, and reliable dissipative functional for
time-dependent current density functional theory.  The local-density
approximation is only a first step. However, once a better
functional is proposed, our formalism provides a simple and elegant
way to test its predictions for the resistance of nanoscale systems.

Finally, we stress that we have focused our attention to the linear-response
regime. It would be interesting and important (although not trivial)
to extend the results presented in this paper to the non-linear
case. Such an extension would allow analysis of the many-body
corrections to the current-voltage characteristics -- and
corresponding dissipation -- of nanoscale systems.

\section{Acknowledgements}
This work has been supported by DOE under Grants No. DE-FG02-05ER46203 and DE-FG02-05ER46204.


\begin{thebibliography}{100}
\bibitem{footnote0}
An additional complication arises when we ask the question of whether the steady-state that interacting electrons establish {\em dynamically} when flowing
in the system is the same as the one imposed {\em a priori} via scattering boundary conditions in a single-particle picture; if that steady state can be
generated with different initial conditions; or if it is unique (see Ref.~\onlinecite{DiVentraTod2004}).
\bibitem{prec1} With abuse of terminology this is often referred to as
``non-equilibrium Green's function formalism''. This is however not
correct since the latter truly refers to a many-body perturbation
technique, while the use of ground-state DFT in extracting
transmission probabilities is done in the context of single-particle
scattering theory.
\bibitem{footnote1}In the absence of a magnetic field, the following stronger conditions are found to hold:\cite{Baranger}
$$
\sum_\alpha \partial_\alpha  \sigma_{\alpha\beta}(\rv,\rv') = 0~,
$$
and
$$
\sum_\beta \partial_\beta^\prime  \sigma_{\alpha\beta}(\rv,\rv')=0~.
$$
\bibitem{footnote2} By contrast, the full current-current response function yields the response of the electric current to the external vector potential only.
The transition from scalar electric potential to a longitudinal
vector potential is done via a gauge transformation.
\bibitem{footnote3} Since the Kohn-Sham system is non-interacting, there is no difference between the (full) current response function $\chi^{(KS)}$ and the proper response function $\tilde \chi^{(KS)}$.
\bibitem{footnote4} Notice that, by virtue of the identities~(\ref{r1}) and~(\ref{r2})   the restricted matrix contains no less information than the original one.
\end{thebibliography}
\end{document}